\documentclass[a4paper,12pt]{article}
\usepackage{url}
\usepackage{graphicx}
\usepackage{color}
\usepackage{ifthen}
\usepackage[greek,english]{babel}

\usepackage{amsmath}
\usepackage{amsthm,amsfonts}
\usepackage{pgfplots}

\usepackage{tikz}
\usepackage[customcolors]{hf-tikz}
\usepgfplotslibrary{colorbrewer}
\usetikzlibrary{mindmap,trees}
\usetikzlibrary{%
  3d,
  arrows,%
  shapes.misc,% wg. rounded rectangle
  shapes.arrows,%
  chains,%
  matrix,%
	fit,%
  positioning,% wg. " of "
  scopes,%
  decorations.pathmorphing,
  shadows,%
  decorations.markings,
  shapes.geometric,
  arrows.meta,bending,
	patterns,
	intersections, 
	pgfplots.fillbetween
}

\usepackage{tikz-3dplot}
\usepackage[pdfusetitle]{hyperref}
\hypersetup{
    colorlinks,
    linkcolor={red!50!black}, 
    citecolor={blue!50!black},
    urlcolor={blue!50!black}
}

\graphicspath{{figures/}{tikz/}}

\newcommand{\gpi}{\textrm{\greektext p}}
\renewcommand{\pi}{\gpi}
\newcommand*{\mat}[1]{\mathbf#1}
\providecommand*{\M}[1]{\mathbf#1}
\providecommand*{\mrm}[1]{\mathrm{#1}}
\providecommand*{\T}[1]{\mathrm{#1}}
\providecommand*{\unit}[1]{\ensuremath{\mrm{\,#1}}}
\renewcommand{\vec}[1]{{\boldsymbol#1}}

\newcommand{\UV}[1]{\hat{\vec{#1}}}
\DeclareMathAccent{\ring}{\mathalpha}{operators}{"17}

\providecommand*{\unit}[1]{\ensuremath{\mrm{\,#1}}}
\providecommand*{\eu}{\ensuremath{\mrm{e}}}

\providecommand*{\ju}{\ensuremath{\mrm{j}}}

\renewcommand{\Re}{\operatorname{Re}}	% The LaTeX standard is not ISO!
\renewcommand{\Im}{\operatorname{Im}}	% The LaTeX standard is not ISO!
\providecommand*{\diff}{\operatorname{d}\!}

\providecommand*{\diffV}{\operatorname{dV}\!}
\newcommand{\norm}[1]{\mathop{\|#1\|}\nolimits}

\newcommand{\eig}{\mathop{\mrm{eig}}}
\newcommand{\atan}{\mathop{\mrm{atan}}}

\newcommand{\qtext}[1]{\quad\text{#1}}

\newcommand{\reg}{\varOmega}

\newcommand{\R}{\mathbb{R}{}}

\newcommand{\C}{\mathbb{C}}	% complex numbers
\providecommand*{\diag}{\operatorname{diag}}
\newcommand{\partder}[2]{\frac{\partial#1}{\partial{#2}}}

\newcommand{\radm}{\varrho} % radiation modes 
\newcommand{\rada}{{\mat{a}}} % radiation mode vector 
\newcommand{\radI}{{\mat{I}}} % radiation mode current vector 
\newcommand{\maxradm}{\bar{\radm}}

\newcommand{\maxrade}{\bar{\nu}}
\newcommand{\Rop}{\mrm{R}} % spherical radial function 
\newcommand{\uop}{\vec{u}} % spherical wave 
\newcommand{\Ds}{\mathcal{D}}
\newcommand{\maxlambda}{\bar{\lambda}}
\newcommand{\minlambda}{\underline{\lambda}}

\newcommand{\maximize}{\mrm{maximize}}

\newcommand{\subto}{\mrm{subject\ to}}

\newcommand{\sphi}{n}

\newcommand{\ie}{\textit{i.e.}\/, }
\newcommand{\eg}{\textit{e.g.}\/, }
\newcommand{\cf}{\textit{cf.}\/, }

\newcommand{\rv}{\vec{r}}
\newcommand{\Ev}{\vec{E}}
\newcommand{\Yv}{\vec{Y}}
\newcommand{\Jv}{\vec{J}}
\newcommand{\Evi}{\vec{E}_{\mrm{i}}}
\newcommand{\kvh}{\hat{\vec{k}}}
\newcommand{\rvh}{\hat{\vec{r}}}
\newcommand{\Fv}{\vec{F}}
\newcommand{\uv}{\vec{u}}
\newcommand{\phivh}{\hat{\vec{\phi}}}
\newcommand{\thetavh}{\hat{\vec{\theta}}}

\newcommand{\Id}{\mat{1}}
\newcommand{\Om}{\mat{0}}
\newcommand{\Xm}{\mat{X}}
\newcommand{\Rm}{\mat{R}}
\newcommand{\Rmr}{\mat{R}_0}
\newcommand{\Xmr}{\mat{X}_0}
\newcommand{\Rml}{\mat{R}_\rho}
\newcommand{\Xml}{\mat{X}_\rho}
\newcommand{\Tm}{\mat{T}}

\newcommand{\Qm}{\mat{Q}}
\newcommand{\Psim}{\mat{\Psi}}
\newcommand{\Lambdam}{\mat{\Lambda}}
\newcommand{\Vm}{\mat{V}}
\newcommand{\Um}{\mat{U}}
\newcommand{\Zm}{\mat{Z}}
\newcommand{\Jm}{\mat{I}}
\newcommand{\fm}{\mat{f}}
\newcommand{\am}{\mat{a}}

\newcommand{\herm}{\mrm{H}}

\newcommand{\basv}{\vec{\psi}}

\newcommand{\perm}{\varepsilon}

\newcommand{\suscd}{\boldsymbol{\chi}}

\newcommand{\rhom}{\boldsymbol{\rho}}
\newcommand{\rhomt}{\widetilde{\boldsymbol{\rho}}}
\newcommand{\rhomr}{\rhom_\mrm{r}}
\newcommand{\rhomi}{\rhom_\mrm{i}}
\newcommand{\rhomrt}{\rhomt_\mrm{r}}
\newcommand{\rhomit}{\rhomt_\mrm{i}}
\newcommand{\rhor}{\rho_\mrm{r}}
\newcommand{\rhoi}{\rho_\mrm{i}}
\newcommand{\rhot}{\tilde{\rho}}
\newcommand{\rhort}{\tilde{\rho}_\mrm{r}}
\newcommand{\rhoit}{\tilde{\rho}_\mrm{i}}
\newcommand{\kr}{\xi} % electrical size
\newcommand{\ka}{\alpha} % 
\newcommand{\Pin}{P_\mrm{in}}

\colorlet{dpurple}{blue!50!red}
\colorlet{dblue}{blue!50!black}
\colorlet{dgreen}{green!50!black}
\colorlet{dred}{red!50!black}
\colorlet{dyellow}{yellow!50!black}
\colorlet{dorange}{orange!50!black}
\definecolor{metal}{RGB}{218,165,32}
\definecolor{diel}{RGB}{1,165,32}
\definecolor{antenna}{RGB}{100,150,162}
\definecolor{breg}{rgb}{0.2,0.6,0.8}%
\definecolor{preg}{rgb}{0.8,0.2,0.2}%
\definecolor{reg}{RGB}{218,165,32}

\pgfplotsset{major grid style={thin,blue!15!white}}
\pgfplotsset{minor grid style={very thin,blue!10!white}}
\tikzset{>=latex}

\begin{document}

\title{Modes, Bounds, and Synthesis of Optimal Electromagnetic Scatterers}

\author{Mats Gustafsson\thanks{M. Gustafsson is with Lund University, Lund, Sweden, mats.gustafsson@eit.lth.se}}

\maketitle

\begin{abstract}
This paper presents an optimal synthesis of material distributions in obstacles for maximal extinction, scattering, or absorption. The material synthesis is based on an explicit construction utilizing the current distribution derived from physical bounds excited from the far-field. The bounds are expressed in radiation modes for materials restricted by their resistivity and characteristic modes for materials restricted by the contrast. The results are valid for arbitrary shapes, and analytical expressions are provided for spherical shapes.
\end{abstract}   

\section{Introduction}
Design of electromagnetic (EM) devices such as antennas, scatterers, absorbers, and filters can be challenging and often requires significant experience and expertise. Traditionally, good designs have relied on physical insight combined with a trial-and-error approach, but computer-based methods utilizing iterative optimization algorithms and inverse design are increasingly aiding in the design process~\cite{Molesky+etal2018,Koziel+Ogurtsov2014,Rahmat-Samii+Michielssen1999}. Despite these advancements, explicit synthesis of shape and material distributions for EM devices remains rare, and achieving optimal synthesis is known only in a few special cases. In this context, a method for the optimal synthesis of material distributions within arbitrary-shaped geometries for maximal scattering and absorption is presented. The technique is valid for arbitrary-shaped regions under far-field illumination and involves synthesizing the material distribution based on the current distribution derived from physical bounds (limits). 

Recently, significant progress has been made in establishing limits for a broad class of antenna and scattering problems valid for arbitrary shapes~\cite{Gustafsson+Nordebo2013,Gustafsson+etal2020,Kuang+Miller2020,Chao+etal2022,Kuang+etal2020}. These limits are derived by transforming the design problem into convex optimization problems over source distributions~\cite{Gustafsson+Nordebo2013}. The performance of realized designs has demonstrated proximity to some of these limits, suggesting practical tightness~\cite{Capek+etal2019b,Gustafsson+etal2009a,Nel+etal2023}. However, in certain cases, designs exhibit performance further from the limits, indicating opportunities for design improvements or the potential to improve the limits, for example, by incorporating additional pointwise or local constraints~\cite{Gustafsson+Capek2020,Kuang+Miller2020}.

In this paper, we extend the limits on the scattering formulation from a fixed illumination scenario~\cite{Gustafsson+etal2020} to optimal far-field excitation. In other words, we determine performance bounds for the combination of material structure and excitation. The power of this illuminating field is fixed, and the optimal performance of material distributions restricted to the design region is determined. It is shown that mode expansions diagonalize the optimization problems and provide explicit representations for several bounds. Moreover, the modes are used to synthesize material distributions in the objects such that they perform according to the bounds, and hence, showing that the bounds are tight. 

Radiation modes~\cite{Gustafsson+etal2020,Schab2016} are used for bounds on cases restricted by material losses, modeled by a minimum resistivity similar to the figure of merit proposed in~\cite{Miller+etal2016}. It is shown that the dominant radiation mode constitute the solution for maximum extinction, scattering, and absorption. Lossless dielectric materials are analyzed using characteristic modes~\cite{Garbacz+Turpin1971,Harrington+Mautz1971,Gustafsson+etal2022a}, where it is shown that the performance is limited by the maximum material contrast.  

This paper is structured as follows: Sec.~\ref{S:Formulation} introduces the problem formulation for optimal material distribution and far-field excitation. Sec.~\ref{S:limits} presents fundamental limits on scattering and absorption constrained by material losses. The general synthesis technique is outlined in Sec.~\ref{S:Synthesis}, while explicit results for spherical structures are detailed in Sec.~\ref{S:SynthesisSph}. Sec.~\ref{S:lossless} delves into limits and explicit synthesis for lossless cases constrained by the material contrast. The paper is concluded in Sec.~\ref{S:Conclusions}.

%%%%%%%%%%%%%%%%%%%%%%%%%%%%%%%%%%%%%%%%%%%%%%%%%%%%%%%%%%%%%
\section{Optimal material and excitation}\label{S:Formulation}
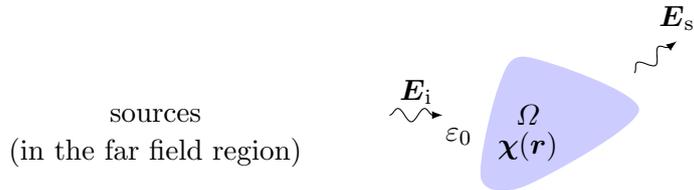
\begin{figure}%
{\centering
\begin{tikzpicture}[scale=0.7,font=\small]
\fill [blue!20!white] plot [smooth cycle,scale=1.1] coordinates {(-0.8,-1.1) (0,-1.2) (1.9,0.1) (0,1) (-0.5,0.5)};    
\node at (0,0) {$\reg$};
\node at (0,-0.6) {$\suscd(\rv)$};
\node at (-1.3,-0.4) {$\perm_0$};
\draw[->,decorate,decoration=snake] (-2.6,0.0) -- node[above] {$\Evi$} +(1.0,0);
\draw[->,decorate,decoration=snake] (2.0,0.8) -- +(0.8,0.6) node[above] {$\Ev_{\mrm{s}}$};
\node at (-7,0) {sources};
\node at (-7,-0.7) {(in the far field region)};
\end{tikzpicture}\par}
\caption{Scattering geometry with an incident electric field $\Evi(\rv)$, scattered field $\Ev_\T{s}(\rv)$, and obstacle $\reg$ with susceptibility $\suscd(\rv)$ in a vacuum background and sources for the incident field far away.}%
\label{fig:scattgeo}%
\end{figure}

Scattered and absorbed powers of an object depend on the illuminating field (excitation), $\Evi$, and materials properties, $\suscd(\rv)$, of the object, see Fig.~\ref{fig:scattgeo}. We assume that the object is confined to a design region, $\reg\subset\R^3$, the incident wave is time harmonic, and the interaction between the object and the illuminating electromagnetic field can be modeled by a complex susceptibility dyadic $\suscd$  or equivalently, complex resistivity dyadic $\rhom=\eta_0\suscd^{-1}/(\ju k)$, where $k$ denotes the free-space wavenumber, $\eta_0$ the intrinsic impedance of free space, and $\ju^2=-1$. 

The aim of this paper is to provide a synthesize technique for the material distribution $\rhom(\rv)$ in the design region $\reg$ such that the resulting scattered or absorbed powers are the maximal possible for the given incident power and material distributions in $\reg$, \ie they reach the physical limit for obstacles confined to this region. To achieve this goal, we first determine fundamental limits on scattered and absorbed powers for obstacles modelled by a resistivity dyadic in a region $\reg$. We assume that the illuminating field is generated by sources far away from the obstacle, \ie can be described by a superposition of plane waves~\cite{Kristensson2016}.

The physical limits are determined by formulating an optimization problem over far-field excitation, $\Evi$ and material distribution $\rhom(\rv)$ in the (design) region $\reg$ of the form 
\begin{equation}
\begin{aligned}
	& \maximize_{\rhom,\Evi} && \T{scattered\ (absorbed)\ power}\\
	& \subto &&  \T{material\ constraints }\\
	& && \T{illumination\ field\ } \Evi \T{\ generated\ far\ away} \\
 & && \T{illuminating\ power\ } P_\T{in}
\end{aligned}  
\label{eq:genopt}
\end{equation}
To formulate and solve these types of optimization problems, we first determine mathematical expressions for the used quantities. The problem is then relaxed to optimization over the induced current density $\Jv$ in $\reg$ which is solved using duality and convex optimization techniques~\cite{Gustafsson+etal2020}.
Solution of the relaxed optimization problem produces an upper bound on the absorbed or scattered power together with a current (source) distribution. These optimal currents are subsequently used to synthesis the material distribution $\rhom(\rv)$.

The illuminating field is described by the electric field $\Evi(\rv)$ assumed to have sources far away from the object, \ie in the far field region, see Fig.~\ref{fig:scattgeo}. This field can be expressed in many ways such as spherical wave expansion~\cite{Hansen1988,Kristensson2016} or plane wave spectra~\cite{Capek+etal2023a} modelled as a superposition of plane waves
\begin{equation}
\Evi(\rv)
=\frac{-\ju k}{4\pi}
\int_{4\pi}\eu^{-\ju k\rv\cdot\kvh}
 \Fv_{\mrm{i}}(\kvh)\diff\Omega_{\kvh}, 
\label{eq:EinPWaves}
\end{equation}
where $\Fv_{\mrm{i}}$ is the (far-field) plane-wave spectrum.
The illuminating field is equivalently described by a superposition of regular spherical waves $\uv_n^{(1)}$~\cite{Hansen1988,Kristensson2016}
\begin{equation}
	\Evi(\rv) = k\sqrt{\eta_0}\sum_{n=1}^M a_n\uv^{(1)}_n(k\rv)
\label{eq:EinSph}
\end{equation}
with the expansion coefficients $a_n$ collected in a column matrix $\M{a}\in\C^{M,1}$. 

The scattered power of an object is determined by integration of the radiated far-field, $\Fv(\rvh)=r\Ev_\T{s}(\rv)\eu^{\ju kr}$ with $\rvh=\rv/r$ and $r=|\rv|$ as $kr\to\infty$, over the unit sphere
\begin{equation}
    P_\T{s} = \frac{1}{2\eta_0}\int_{4\pi} |\Fv(\rvh)|^2\diff\Omega = \frac{1}{2}|\M{f}|^2
\end{equation}
or equivalently by summation of the power of each radiated spherical wave coefficient $f_n$ collected into the column matrix $\fm$. The incident (excitation) power is similarly determined by integration of the plane wave spectrum~\eqref{eq:EinPWaves} or summing spherical wave coefficients $\am$
\begin{equation}
    P_{\T{in}} = \frac{1}{2\eta_0}\int_{4\pi} |\Fv_\T{i}(\rvh)|^2\diff\Omega
%    =\frac{1}{2}|\fm_\T{i}|^2
    =\frac{1}{8}|\am|^2.    
    \label{eq:Pin}
\end{equation}
The factor $1/8$ stems from the expansion of the incident field in regular waves (Bessel functions), which can be seen as a sum of incident and outgoing waves (Hankel functions)~\cite{Kristensson2016}.

Material losses are related to the real part of the complex resistivity dyadic $\rhom$~\cite{Gustafsson+etal2020} with the absorbed power in an object given by 
\begin{equation}
P_{\mrm{a}}=\frac{\Re}{2}\int_\reg\Jv^{\ast}(\rv)\cdot\rhom(\rv)\cdot\Jv(\rv)\diffV
 =\frac{1}{2}\Jm^{\herm}\Rml\Jm.
\label{eq:Pa}
\end{equation}
The current density $\Jv$ is expanded in basis functions $\basv_n(\rv)$ as $\Jv(\rv)\approx\sum_{n=1}^P I_n\basv_n(\rv)$ to express the absorbed power in matrix form by collecting the expansion coefficients $I_n$ in a column matrix $\Jm$. In~\eqref{eq:Pa}, the superscript ${}^\herm$ denotes Hermitian transpose.

The resistivity dyadic $\rhom$ has dimension $\unit{\Omega m}$ and it is convenient to introduce a dimensionless resistivity dyadic $\rhomt=k\rhom/\eta_0$ which has a simple inverse relation with the susceptibility dyadic
\begin{equation}
	\suscd = \frac{\eta_0}{\ju k}\rhom^{-1}=-\ju\rhomt^{-1}.
\label{eq:chi2rho}
\end{equation}
The real part of $\rhomt=\rhomrt+\ju\rhomit$ is identified with the material figure of merit (FoM) proposed in~\cite{Miller+etal2016}, \eg expressed for an isotropic resistivity as $\rhort = -\Im\chi/|\chi|^2$. 
The corresponding imaginary part (reactivity) $\Im\{\rhom\}$ is arbitrary, which can be interpreted as a freedom to tune the reactivity of the obstacle for fixed $\rhomr$.  

For simplicity, in this paper we follow~\eqref{eq:EinSph} by expressing the incident and radiated fields in spherical waves. 
The scattered (radiated) field is also expressed in the induced currents as
\begin{equation}
    P_\T{s} = \frac{1}{2}|\fm|^2 
    = \frac{1}{2}\Jm^{\herm}\Um^{\herm}\Um\Jm
    =\frac{1}{2}\Jm^{\herm}\Rmr\Jm,
    \label{eq:Ps}
\end{equation}
where $\M{U}$ denotes the projection matrix of the regular spherical waves onto the basis functions~\cite{Gustafsson+etal2015b}, see App.~\ref{S:SphWaves}. The extincted (or total) power is the sum of the absorbed~\eqref{eq:Pa} and scattered~\eqref{eq:Ps} powers
\begin{equation}
    P_\T{t} = P_\T{a}+P_\T{s}
    =\frac{1}{2}\Re\{\Jm^{\herm}\Vm\},
    \label{eq:Pt}
\end{equation}
which can be expressed as a product between the current and the excitation $\Vm=\Um^{\herm}\am$.

%%%%%%%%%%%%%%%%%%%%%%%%%%%%%%%%%%%%%%%%%%%%
\section{Physical limits}\label{S:limits}
The scattering model, based on the Method of Moments (MoM), is expressed as $\Zm\Jm=\Vm$, where $\Zm\in\C^{P,P}$ represents the system matrix employing $P$ basis functions~\cite{Harrington1968}. We assume that a sufficient number of basis functions are used such that modelling errors can be neglected. 
The MoM equation is relaxed to $\Jm^{\herm}\Zm\Jm=\Jm^{\herm}\Vm$, which can be interpreted as conservation of (complex) power or allowing for a material composed by a mixture between the considered material ($I_n\neq 0$) and the background material ($I_n=0$)~\cite{Gustafsson+etal2020}. 

Considering first maximization of the extincted power~\eqref{eq:Pt} over material distributions within $\reg$ and excitations $\am$ leading to the optimization problem
\begin{equation}
\begin{aligned}
    & \mathop{\mrm{maximize}}_{\Jm\in\C^{P,1},\am\in\C^{M,1}} && P_\T{tZ}=\frac{1}{2}\Re\{\Jm^{\herm}\Vm\}\\
    & \subto && \Jm^{\herm}\Zm\Jm=\Jm^{\herm}\Vm \\
		& && \Vm = \Um^{\herm}\am \\
		& && |\am|^2 = 8\Pin,
        %\qtext{with } \Jm\in\C^{P,1}
\end{aligned}  
\label{eq:MaxPtZ}
\end{equation}
where the used constraints describe the relaxed MoM equation, far-field excitation expressed in spherical waves, and the fixed incident power~\eqref{eq:Pin}.

Solution of this optimization problem~\eqref{eq:MaxPtZ} determines an upper limit on the extincted power achievable by objects confined within a designated design region $\reg$, synthesized from either a material $\rhom(\rv)$ or free space (or background material) from a far-field excitation with incident power $\Pin$.

\subsection{Limits for material with minimal resistivity}
Performance in terms of extinction, scattering, and absorption for objects confined to a region $\reg$ and synthesized with materials modeled by a complex resistivity $\rhom(\rv)$, with a real value such that $\Re\{\rhom(\rv)\}\succeq\rhomr(\rv)$, \ie the material loss component is bounded from below and the imaginary part (reactivity) $\Im\{\rhom(\rv)\}$ is arbitrary. The material inequality is interpreted as allowing materials with higher resistivity (more lossy) than the prescribed value. This is also consistent with mixing between a material and a lossless background, which generally produces a homogenized material with higher resistivity~\cite{Milton2002}. 

We relax the optimization problem~\eqref{eq:MaxPtZ} to materials exhibiting a minimal resistivity. This relaxation leads to the condition $\Re\{\Jm^{\herm}\Zm\Jm\}\geq\Jm^{\herm}(\Rmr+\Rml)\Jm=\Jm^{\herm}\Rm\Jm$, where $\Rml$ is derived from $\rhomr(\rv)$ using~\eqref{eq:Pa} and inserted into~\eqref{eq:MaxPtZ}, where also the constraint on $\Im\{\Zm\}$ is dropped. The upper limit on extinction for minimum resistivity $\Re\{\rhom(\rv)\}\succeq\rhomr(\rv)$ and incident power $\Pin$ is determined through the optimization problem
\begin{equation}
\begin{aligned}
	& \maximize_{\am,\Jm} && P_\T{tR}=\frac{1}{2}\Re\{\Jm^{\herm}\Um^{\herm}\am\}\\
	& \subto &&  \Jm^{\herm}\Rm\Jm \leq \Re\{\Jm^{\herm}\Um^{\herm}\am\}\\
	& && |\am|^2 = 8\Pin.
\end{aligned}  
\label{eq:maxPtR}
\end{equation}
This is a QCQP with two constraints, which can be solved in two steps by first optimizing over currents $\Jm$ using Lagrange duality~\cite{Boyd+Vandenberghe2004} and secondly over excitations $\am$ resulting in an eigenvalue problem.

The first step starts with maximization over the currents $\Jm$ for an arbitrary fixed illumination $\am$ using Lagrangian duality~\cite{Boyd+Vandenberghe2004,Gustafsson+etal2020} resulting in
\begin{equation}
\begin{aligned}
	& \maximize_{\am} && P_\T{tR}=\frac{1}{2}\am^{\herm}\Um\Rm^{-1}\Um^{\herm}\am\\
	& \subto && |\am|^2 = 8\Pin.
\end{aligned}
\label{eq:maxPtRa}
\end{equation}
The second step with optimizing over $\am$ is recognized as a Rayleigh quotient~\cite{Harrington1968} and is solved by the maximal eigenvalue of $\Um\Rm^{-1}\Um^{\herm}$ giving the fundamental limit 
\begin{equation}
\frac{P_\T{t}}{\Pin}\leq \frac{P_\T{tR}}{\Pin} 
=4\max\eig(\Um\Rm^{-1}\Um^{\herm}).
%\leq 4
\label{eq:maxPtReig}
\end{equation}
This bound on the extincted power is similar to the formulation for optimal illumination in~\cite{Kuang+Miller2020}.

The upper bound on the normalized extincted power, $P_\T{t}/\Pin$, is given by the maximal eigenvalue $\maxrade=\max_n\{\nu_n\}$ of $\Um\Rm^{-1}\Um^{\herm}$, which can alternatively be written 
\begin{equation} 
\Um\Rm^{-1}\Um^{\herm}\rada_n = \nu_n\rada_n 
	% \Rightarrow
	% \Um^{\herm}\Um\Rm^{-1}\Um^{\herm}\rada_n = \nu_n\Um^{\herm}\rada_n\\
	\Rightarrow
    \Rmr\radI_n = \radm_n\Rml\radI_n,	
\label{eq:RadmEig}
\end{equation} 
where $\Rmr=\Um^{\herm}\Um$, $\radI_n=\Rm^{-1}\Um^{\herm}\rada_n$, and $\nu_n=\radm_n/(1+\radm_n)$ are used. 
These modes are recognized as radiation modes~\cite{Schab2016,Gustafsson+etal2020} and associated with
the most efficient radiating current distribution determined by maximizing the Rayleigh quotient~\cite{Harrington1968} 
\begin{equation}
	\max_{\Jm}\frac{P_{\T{rad}}}{P_{\T{loss}}}
	=\max_\Jm\frac{\Jm^{\herm}\Rmr\Jm}{\Jm^{\herm}\Rml\Jm}
	=\max_n\radm_n = \maxradm.
\label{eq:RadmRayleigh}
\end{equation}
Radiation mode currents $\radI_n$ are real valued %and excitations $\rada_n$ 
and orthogonal over the material resistivity $\radI_m^{\herm}\Rml\radI_n = \delta_{mn}$, \cf~\eqref{eq:Pa} and far field $\radI_m^{\herm}\Rmr\radI_n = \radm_n\delta_{mn}$, \cf~\eqref{eq:Ps}. 
They constitute a complete basis to expand currents and far fields. Radiation modes have a close connection with maximal radiation efficiency and gain of antennas~\cite{Gustafsson+Capek2019} and the number of degrees of freedom~\cite{Ehrenborg+Gustafsson2020,Gustafsson2024a}. 

\begin{figure}
    \centering
    \includegraphics[width=0.65\textwidth]{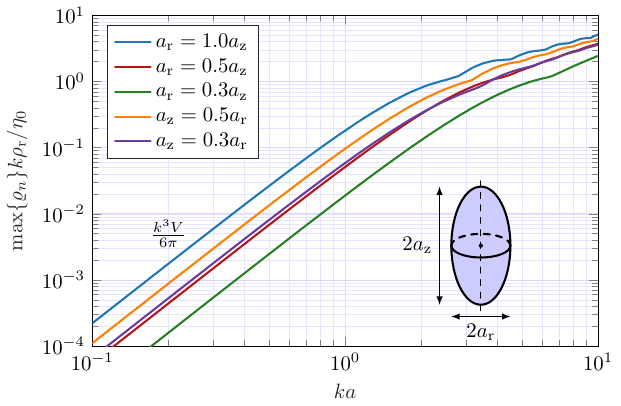}
    \caption{Normalized radiation modes $\maxradm=\max\{\radm_n\}$ for spheroidal regions with $a=\max\{a_\T{r},a_\T{z}\}$. The low-frequency asymptotic is proportional to the volume $V=|\reg|$~\cite{Gustafsson+etal2020}.}
    \label{fig:radmSphoids}
\end{figure}

The maximal radiation modes $\maxradm$ normalized by the resistivity for homogeneous spheroids with semi-axis $a_\T{r}$ and $a_\T{z}$ are depicted in Fig.~\ref{fig:radmSphoids}. The normalized radiation modes are proportional to the electric volume $k^3V/(6\pi)$ for electrically small sizes~\cite{Gustafsson+etal2020}. Their behavior is more complex for larger sizes and \eg the dominant mode $\maxradm$ for the spherical case $a_\T{r}=a_\T{z}$ switches between TM and TE modes as indicated by the oscillations in the figure.

Radiation modes diagonalize the optimization problem~\eqref{eq:maxPtR} expressing the bound~\eqref{eq:maxPtReig} solely in the dominant radiation mode 
\begin{equation}
\frac{P_\T{t}}{\Pin}\leq \frac{P_\T{tR}}{\Pin} 
=\frac{4\maxradm}{1+\maxradm}.
%\leq 4
\label{eq:maxPtRradm}
\end{equation}
Radiation modes are also used for the corresponding bounds on scattering and absorption. 

\subsection{Scattering and absorption}
The maximum scattering for an object confined to a region $\reg$, constructed of materials with resistivity $\Re\{\rhom\}\succeq\rhomr$, and illuminated by power $P_{\mrm{in}}$ is determined from the optimization problem
\begin{equation}
\begin{aligned}
	& \maximize_{\am,\Jm} && P_{\mrm{sR}}=\frac{1}{2}\Jm^{\herm}\Rm_0\Jm\\
	& \subto &&  \Jm^{\herm}(\Rml+\Rmr)\Jm \leq \Re\{\Jm^{\herm}\Um^{\herm}\am\}\\
	& && |\am|^2 = 8P_{\mrm{in}}.
\end{aligned}  
\label{eq:PsR}
\end{equation}
This problem is solved using Lagrangian duality for optimization over the current $\Jm$ and a fixed excitation $\am$~\cite{Gustafsson+etal2020}
\begin{equation}
\begin{aligned}
	& \min_{\upsilon\geq \maxrade}\max_{\am} &&  P_{\mrm{sR}}=\frac{1}{8}\upsilon^2\am^{\herm}\big(\Um((\upsilon-1)\Rm_0+\upsilon\Rml)^{-1}\Um^{\herm}\big)\am\\
	& \subto &&  |\am|^2 = 8P_{\mrm{in}},
\end{aligned}  
\label{eq:PsRd}
\end{equation}
where $\maxrade=\maxradm/(1+\maxradm)$ denotes the efficiency of the largest radiation mode $\maxradm$. Maximization over $\M{a}$ and diagonalizing using radiation modes~\eqref{eq:RadmEig} results in
\begin{multline}
	\frac{P_{\mrm{sR}}}{P_{\mrm{in}}} = \min_{\upsilon\geq \maxrade} \upsilon^2\max\eig\big(\Um((\upsilon-1)\Rm_0+\upsilon\Rml)^{-1}\Um^{\herm}\big)\\
	=\min_{\upsilon\geq \maxrade}\max_n \frac{\upsilon^2\radm_n}{\upsilon-\radm_n+\radm_n\upsilon}
	=\min_{\upsilon\geq \maxrade} \frac{\upsilon^2\maxradm}{\upsilon-\maxradm+\maxradm\upsilon},
 \label{eq:PsRdual}
\end{multline}
which is monotonically increasing with $\radm_n$ for all $\upsilon$ implying maximization over $n$ by $\maxradm$. Minimized over $\upsilon$ from the stationary point  $\upsilon=2\maxradm/(1+\maxradm)$ producing an upper bound on the normalized scattered power~\eqref{eq:Ps}
\begin{equation}
	\max_{\am}\frac{P_{\mrm{sR}}}{P_{\mrm{in}}} 
	=\frac{4\maxradm^2}{(1+\maxradm)^2}.
\label{eq:PsRmode}
\end{equation}
The solution shows that the same excitation and dominant radiation mode $\maxradm$ maximizes extinction and scattering. 
The normalized scattering approaches 4 from below as $\maxradm\to\infty$, \eg in the limit of lossless materials $\rhor\to 0$. 

Maximum absorption is formulated similarly to the extinction and scattering problems and given by the optimization problem
\begin{equation}
\begin{aligned}
	& \maximize_{\am,\Jm} && P_{\mrm{aR}}=\frac{1}{2}\Jm^{\herm}\Rml\Jm\\
	& \subto &&  \Jm^{\herm}(\Rml+\Rmr)\Jm \leq \Re\{\Jm^{\herm}\Um^{\herm}\am\}\\
	& && |\am|^2 = 8P_{\mrm{in}}
\end{aligned}  
\label{eq:PaRopt}
\end{equation}
solved using Lagrangian duality and radiation modes~\cite{Gustafsson+etal2020}
\begin{equation}
	\max\frac{P_{\mrm{aR}}}{P_{\mrm{in}}} 
	=\min_{\upsilon\geq 1}\max_n \frac{\upsilon^2\radm_n}{\radm_n\upsilon+\upsilon-1}
\label{eq:PaRdual}
\end{equation}
which is monotonically increasing in $\radm_n$ for $\upsilon>1$ and has a stationary point at $\upsilon=2/(1+\maxradm)$. The stationary point is in the region $\upsilon\geq 1$ if $\maxradm\in[0,1]$ producing the bound  
\begin{equation}
	\max\frac{P_{\mrm{aR}}}{P_{\mrm{in}}}
	=\begin{cases}
		\dfrac{4\maxradm}{(1+\maxradm)^2}, & \maxradm\in[0,1] \\
		\hfill 1\hfill, & \maxradm\geq 1.	
	\end{cases}
\label{eq:PaRmode}
\end{equation}
The solution for $\maxradm < 1$ utilizes the dominant radiation mode and is similar to the solutions for maximal extinction~\eqref{eq:maxPtRradm} and scattering~\eqref{eq:PsRmode}. The case with $\maxradm > 1$ can be interpreted as resulting from an inactive first constraint in~\eqref{eq:PaRopt}, leading to total absorption of the available illuminating power. Illumination and synthesis in this scenario are generally non-unique and can be derived from any radiation mode with $\radm_n \geq 1$.

\begin{figure}
{\centering
\begin{tikzpicture}[scale=1,font=\small]
  \begin{axis}[
width=0.45\textwidth,
height=0.35\textwidth,
at={(0cm,0cm)},
scale only axis,	
grid=both,
ymin=0.01, ymax=4,
xmin=0.01, xmax=100, 
xtick={0.01,0.1,1,10,100},
xmode=log,
ymode=log,
log ticks with fixed point,
xlabel={maximal radiation mode $\maxradm=\max \radm_n$},
ylabel=$P/P_{\mrm{in}}$]
\addplot[very thick,dblue,domain=0.01:100]  {4*\x/(1+\x)};
\node[dblue] at (axis cs:0.9,3) {$P_\mrm{t}$};
\addplot[very thick,dgreen,domain=0.01:100]  {4*\x*\x/(1+\x)/(1+\x)};
\node[dgreen] at (axis cs:0.5,0.14) {$P_\mrm{s}$};
\node[dblue] at (axis cs:45,3) {$\approx 4$};
\node[dgreen] at (axis cs:0.2,0.02) {$\approx 4\maxradm^2$};
\addplot[very thick,dred,domain=0.01:1]  {4*\x/(1+\x)/(1+\x)};
\draw[very thick,dred] (axis cs:1,1) -- (axis cs:100,1);
\node[dred] at (axis cs:3,0.7) {$P_\mrm{a}$};
\node[dred] at (axis cs:45,0.7) {$\approx 1$};
\node[dred] at (axis cs:0.03,0.05) {$\approx 4\maxradm$};
\end{axis}
%\end{tikzpicture}
%\begin{tikzpicture}[scale=1.0]
\begin{scope}[xshift=10.5cm,yshift=2.3cm]    
\def\rad{3.5cm}
\begin{scope}
\clip (-\rad,-0.6*\rad) rectangle (0,0.6*\rad);
\foreach \x in {1,...,5}{  	
\draw[densely dashed] (0,0) circle (0.2*\x*\rad);
};
\end{scope}
\begin{scope}
    \clip (-0.5*\rad,0) circle (0.5*\rad);
\foreach \x in {40,...,1}{  	
		\pgfmathsetmacro\k{\x*2}
	  \fill[blue!\k] (0,0) circle (0.025*\x*\rad);
};		
\foreach \x in {5,...,1}{  	
	  \draw[black] (0,0) circle (0.2*\x*\rad);
};		
\end{scope}
\draw[thick] (-0.5*\rad,0) circle (0.5*\rad);
	\draw[->] (-1.1*\rad,0) -- (0.9,0) node[below] {$\mathrm{Re}\{t_n\}$};
	\draw[->] (0,-0.6*\rad) -- (0,0.6*\rad) node[right] {$\mathrm{Im}\{t_n\}$};
	\node[below,red!50!black] at (-\rad,0) {$-1$};
	\fill[red!50!black] (-0.5*\rad,0) circle (0.02*\rad);
	\node[below,red!50!black] at (-0.5*\rad,0) {$-0.5$};

\foreach \x in {1,...,2}{  	
\pgfmathsetmacro\k{\x*0.2}
\node[right] at (0,0.2*\x*\rad) {$\k$};
};
\end{scope}
\end{tikzpicture}

\par}	

\caption{(left) Upper bounds on the extincted $P_\T{t}$, scattered $P_\T{s}$, and absorbed $P_\T{a}$ powers as function of the maximal radiation mode $\maxradm=\max\{\radm_n\}$. (right) Corresponding bounds expressed for transition matrix eigenvalues, $t_n$, illustrated for $\maxrade\in\{0.2,0.4,0.6,0.8,1\}$.}
    \label{fig:MaxTotScattAbsP}
\end{figure}
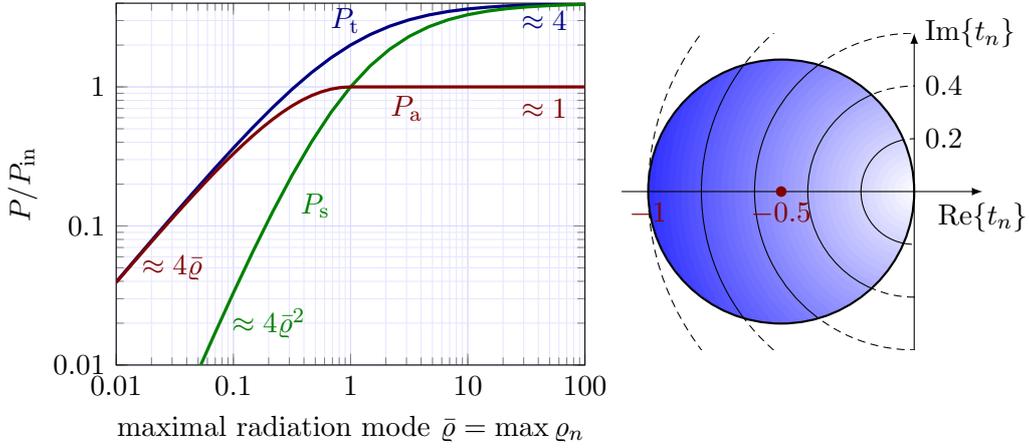

Figure~\ref{fig:MaxTotScattAbsP} illustrates the fundamental limits on normalized extinct, scattered, and absorbed powers as a function of the dominant radiation mode $\maxradm$. The absorbed power $P_\T{a}$ is constrained by $P_\T{in}$ due to power conservation, as the object cannot absorb more power than is incident upon it. The extincted $P_\T{t}$ and scattered $P_\T{s}$ powers are capped at $4P_\T{in}$, which does not violate power conservation because phase shifts in the scattered field can enhance scattering, as \eg noted in the extinction paradox (shadow scattering)~\cite{Peierls1979}. Scattering and absorption limits are equal when $\maxradm = 1$ and decay as $4\maxradm$ for absorption and $4\maxradm^{2}$ for scattering as $\maxradm \to 0$.

The bounds on extinct, scattered, and absorbed powers correspond to bounds on eigenvalues $t_n$ of transition matrices $\M{T}$ for objects composed of materials with $\Re\{\rhom\} \geq \rhomr$. The bound on normalized extinct~\eqref{eq:maxPtRradm} and scattered~\eqref{eq:PsRmode} powers translates to 
\begin{equation}
    -\Re t_n\leq \frac{\maxradm}{1+\maxradm}
    =\maxrade
    \quad\text{and }
    |t_n|\leq \maxrade,%\frac{\maxradm}{(1+\maxradm)}
    \label{eq:ExtSTbound}
\end{equation}
respectively. Limits on absorption are easier expressed using the eigenvalues $s_n=2t_n+1$ of the scattering matrix $\M{S}=2\M{T}+\Id$  
\begin{equation}
    1-|s_n|^2
    =-4t_n-4|t_n|^2
    =1-4\left|t_n+\frac{1}{2}\right|^2
    \leq 
    \begin{cases}
		\dfrac{4\maxradm}{(1+\maxradm)^2}, & \maxradm\in[0,1] \\
		\hfill 1\hfill, & \maxradm\geq 1.	
\end{cases}
\label{eq:AbsTbound}
\end{equation}
Limits on the transition matrix eigenvalues $t_n$ are illustrated to the right in Fig.~\ref{fig:MaxTotScattAbsP}. The range of $t_n$ lies within a circular region with a radius of $0.5$ centered at $-0.5$ in the complex plane~\cite{Kristensson2016}. When constrained by radiation modes with maximal efficiency $\maxrade$, this range is further restricted to the intersection of this circular region with another circular region of radius $\maxrade$ centered at $0$, as shown in Fig.~\ref{fig:MaxTotScattAbsP}
This perspective unifies the limits on scattering, extinction, and absorption. The radius of the circular region centered at $0$ is determined by the scattering bound $|t_n| \leq \maxrade$, the extinction bound from the intersection with the real axis, and the absorption bound depends on the distance to the point $-0.5$. The region contains the point $-0.5$ for $\maxrade \geq 0.5$, in accordance with the value of 1 in~\eqref{eq:AbsTbound}.

The limits presented in Fig.~\ref{fig:MaxTotScattAbsP} apply to arbitrarily shaped objects composed of materials with $\Re\{\rhom\} \geq \rhomr$. The imaginary part, $\Im\{\rhom(\rv)\}$, does not factor into the optimization problems~\eqref{eq:maxPtRradm},~\eqref{eq:PsRmode}, and~\eqref{eq:PaRmode}. This reactivity serves as a free dyadic function that can be used to tune the object to resonance. The physical bounds are tight if there exists a reactivity $\Im\{\rhom(\rv)\}$ such that the extinct, scattered, or absorbed powers of an object made with $\rhom = \rhomr + \ju\rhomi$ match the established limits. An explicit material synthesis based on the currents $\M{I}$ computed from~\eqref{eq:maxPtR},~\eqref{eq:PsR}, and~\eqref{eq:PaRopt} is presented next.

%%%%%%%%%%%%%%%%%%%%%%%%%%%%%%%%%%%%%%%%%%%%%%%%%%%%%%%%%%%%%%
\section{Material synthesis}\label{S:Synthesis}
The operator $\Um\Rm^{-1}\Um^{\herm}$, utilized for determining limits on the extincted power~\eqref{eq:maxPtReig}, resembles the evaluation of the transition matrix $\Tm=-\Um\Zm^{-1}\Um^{\herm}$ from a MoM matrix $\Zm=\Rm+\ju\Xm$~\cite{Gustafsson+etal2022a}. The upper bound can thus be interpreted as an ideal material that annuls the reactive (imaginary) part in the MoM system $\Xm=\M{0}$. The reactance matrix $\Xm=\Xm_0+\Xml$ comprises a free-space part $\Xm_0$ and a material part $\Xml$~\cite{Gustafsson+etal2020}. The free space part $\Xmr$ is non-local, seemingly necessitating a non-diagonal material matrix $\Xml=-\Xmr$, corresponding to non-local material interaction. However, it turns out to be sufficient to eliminate the reactance for the particular excitation $\Vm=\Um^{\herm}\am$ from the solutions $\am$ of~\eqref{eq:maxPtRradm},~\eqref{eq:PsRmode}, or~\eqref{eq:PaRmode}.

The synthesis technique is based on showing that the material reactance matrix $\Xml$ and imaginary resistance $\Im\{\rhom(\rv)\}=\rhomi(\rv)$ can be synthesized to produce arbitrary radiation modes~\eqref{eq:RadmEig} and hence optimal performance according to~\eqref{eq:maxPtReig},~\eqref{eq:PsRmode}, or~\eqref{eq:PaRmode}. 

Synthesis of the material reactivity $\rhomi(\rv)$ for a given real valued current $\Jm_n$ and excitation $\Vm_n=\Rm\Jm_n$ is based on subtracting $\Zm\Jm_n=\Vm_n$ and $\Rm\Jm_n=\Vm_n$ which reduces to 
\begin{equation}
	\Xml\Jm_n = -\Xm_0\Jm_n.
\label{eq:Xsynthesis}
\end{equation} 
For simplicity using a volumetric MoM formulation with piece-wise constant basis functions~\cite{Polimeridis+etal2014} resulting in a diagonal (real-valued) material matrix $\Xml$. This reduces the relation~\eqref{eq:Xsynthesis} assuming non-vanishing currents to $P$ constraints for $P$ unknowns according to $X_{\rho,nn}I_n=-\sum X_{0nm}I_m$ for $n=1,...,P$. 
Solving this problem by dividing with $I_n$ produces a unique material reactance matrix $\Xml$ according to
\begin{equation}
	\Xml = -\diag(\Xm_0\Jm_n  \oslash \Jm_n), 
\label{eq:PtRF2X}
\end{equation}
where $\oslash$ denotes the element-wise (Hadamard) division. 
Here, we note that division with zero for some element produces an infinite reactivity which translates to free space $\chi=0$ at that element position and direction.  
For numerical evaluation a minimization $\min\norm{\Xml\Jm_n-\Xm_0\Jm_n}$  with some regularization, \eg isotropic, minimum variation, can also be used.

This simplified derivation is based on the assumption of non-zero current elements which generally does not hold for radiation modes, such as the three orthogonal modes present in electrically small volumes (e.g., in a cartesian coordinate system). Vanishing components are addressed in various ways. One approach involves discarding the design degrees of freedom associated with zero elements in $\Jm$, effectively replacing the material with its background, such as vacuum. This synthesis technique yields a reactance matrix that forms an obstacle satisfying $\Zm\Jm=\Vm$, thereby achieving performance according to fundamental limits. A drawback of this approach is the potential for strong anisotropy in the synthesized material.

An alternative technique is to form a linear combination between the dominant radiation modes (typically 3 for volumes~\cite{Gustafsson+etal2020}) producing a material with less anisotropy. Particularly in the electrically small region, the synthesized material becomes isotropic with performance according to the fundamental limits. These possibilities also highlight that the optimal material distribution is not unique.

The outlined synthesis applies for maximal extinction and scattering. It also applies for maximal absorption for cases with maximal radiation mode $\maxradm\leq 1$. For modes with $\radm_n\geq 1$, the material distribution can \eg be synthesized by increasing the resistivity until the first constraint becomes equality and subsequently using the procedure in~\eqref{eq:PtRF2X}. The synthesized structure can be interpreted as producing a real-valued transition matrix eigenvalue $t_n$ at $k_0a$ with equality in~\eqref{eq:ExtSTbound} or~\eqref{eq:AbsTbound} as shown in Fig.~\ref{fig:MaxTotScattAbsP}.

%%%%%%%%%%%%%%%%%%%%%%%%%%%%%%%%%%%%%%%%%%%%%%%%%%%%%%%%
\section{Material synthesis for spherical regions}\label{S:SynthesisSph}
Material distributions linked to radiation modes can be analytically synthesized for spherical regions with a homogeneous isotropic resistivity $\rhor$. This derivation relies on the decomposition of the Green's dyadic in spherical waves, as outlined in~\ref{S:SphWaves}. Radiation modes for a spherical region with homogeneous isotropic resistivity are proportional to regular spherical waves~\cite{Gustafsson+etal2020}. 

\begin{figure}
    \centering
    \includegraphics[width=0.7\textwidth]{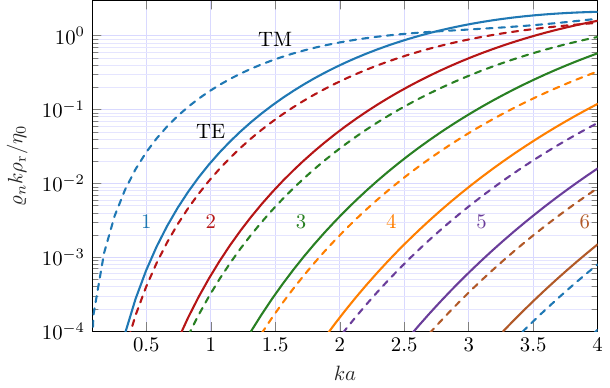}
    \caption{Normalized radiation modes for a spherical region with radius $a$ and resistivity $\rhor$. TM modes showed by dashed curves and TE modes by solid curves for orders $l=1,2,\ldots$.}
    \label{fig:radmodessphereTMTE}
\end{figure}

Normalized radiation modes for spherical regions are depicted in Fig.~\ref{fig:radmodessphereTMTE}. The radiation modes are closely linked to the spherical waves and decomposed into TM and TE modes of order $l=1,2,\ldots$. The modes are degenerate with a Fourier type expansion in the azimuthal direction. The electric dipole ($l=1$) modes (TM) dominates for electrically small sizes and up to $ka\approx 2.74$ after which the TE dipole takes over. Quadruple ($l=2$) and higher order modes are weaker for electrically small object but become large as $ka$ increases.

\begin{figure}
    \centering
    \includegraphics[width=0.25\textwidth]{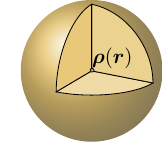}
    \begin{tikzpicture}
    \def\a{15mm}
    \def\b{10mm}
    \fill[metal!70!white] (0,0) circle(\a);
    \draw[<->] (0,0) -- node[above] {$a$} (-1.2,0.9);
    \draw[->] (0,0) -- node[above] {$\rv$} (0.8,0.6);
    \draw[->] (0.8,0.6) -- node[above] {$\rvh$} +(0.6,0.45);
    \draw[->] (0.8,0.6) -- node[right] {$\thetavh$} +(0.45,-0.6);    
    \node at (0,-0.55) {$\rhom(\rv)$};    
    \end{tikzpicture}
    \caption{Spherical design region with an anisotropic resistivity $\rhom(\rv)=\rho_\T{rr}(r)\rvh\rvh+\rho_\T{tt}(r)(\thetavh\thetavh+\phivh\phivh)$ expressed in a spherical coordinate system with radial $\rvh$, polar $\thetavh$, and azimuthal $\phivh$ components.}
    \label{fig:Sphere3D}
\end{figure}

Explicit synthesis of $\Im\{\rhom\}$ results in isotropic materials for TE modes and anisotropic materials for TM modes expressed in spherical coordinates as illustrated in Fig.~\ref{fig:Sphere3D}. The material distributions from~\eqref{eq:PtRF2X} have closed form expressions in spherical Bessel function, see App.~\ref{S:SphWaves}.

\begin{figure}%
{\centering
\includegraphics[width=0.49\textwidth]{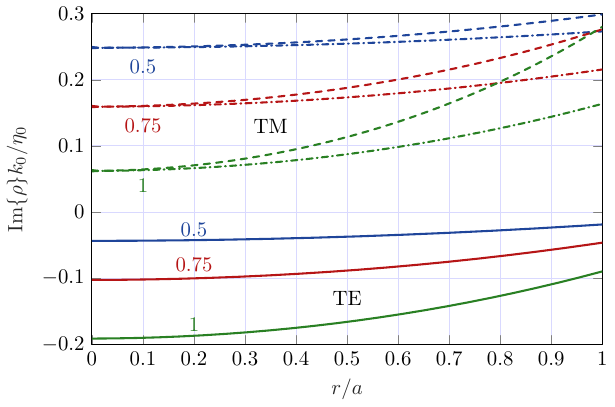}%
\includegraphics[width=0.49\textwidth]{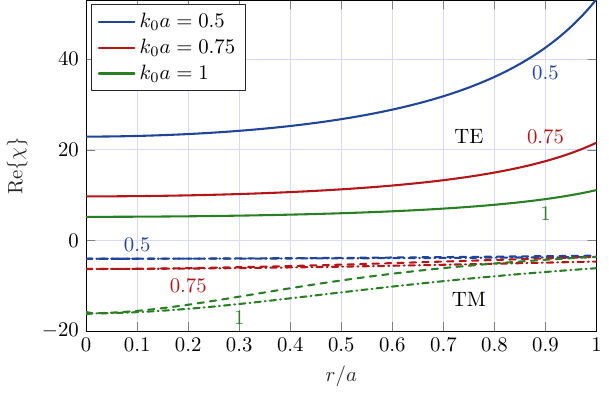}
\par}
\caption{Synthesis of optimal material distributions in a spherical region for dominant TM (dashed for $\thetavh\thetavh,\phivh\phivh$ and dashed dotted for $\rvh\rvh$ components) and TE modes (solid) evaluated at $k_0a\in\{0.5,0.75,1\}$. (left) normalized reactivity $\Im\rhomt(\rv)$ and (right) susceptibility $\Re\{\suscd(r)\}$ determined for negligible losses $\rhort=10^{-4}$.}%
\label{fig:SynthesisSph}%
\end{figure}

The closed form expressions for arbitrary modes are lengthy so starting with the magnetic (TE) dipole mode. Synthesis yields an isotropic material, $\rhom(\rv)=\rho(r)\Id$, where $\Id$ denotes the identity dyadic, with normalized reactivity
\begin{equation}
\Im\{\rhot(r)\}
=1+ \frac{\kr^2\tan\kr}{2(\kr-\tan\kr)}
+\frac{\sin(2\ka)}{2\ka}-\frac{\cos^2\ka}{2},
\label{eq:rhoiTE1full}
\end{equation}
where $\kr=k_0r$ and $\ka=k_0a$ are used. The reactivity is depicted in Fig.~\ref{fig:SynthesisSph} for the cases $k_0a\in\{0.5,0.75,1\}$ together with the corresponding susceptibility evaluated for $\rhort=10^{-4}$. The reactivity is negative up to $k_0a\approx 2.8$ (first zero of $\cos\ka=-\ka\sin\ka$) corresponding to a positive susceptibility, see Fig.~\ref{fig:SynthesisSph}. 

The reactivity approaches zero for small sizes corresponding to a high contrast material, \ie a large susceptibility. Asymptotic expansions for electrically small sizes (long wavelength) are simpler to interpret
\begin{equation}
\Im\{\rhot(r)\} 
\approx -\frac{\ka^2}{6}
-\frac{\ka^4}{30}
+\frac{\ka^6}{105}
+\frac{\kr^2}{10} 
+\frac{\kr^4}{350}
.
\label{eq:rhoTEsph}
\end{equation}
This asymptotic expansion has negligible errors (less than a relative error of $1\%$ for $k_0a=1$) for the considered cases in Fig.~\ref{fig:SynthesisSph}. 

Radiation modes for the magnetic dipole mode (TE) are weak for small electrical sizes $ka\ll 1$ but starts to dominate over the TM modes around $k_0a\approx 2.74$, see Fig~\ref{fig:radmodessphereTMTE}. 
Although the response of TE modes is weaker, it is generally realizable by ordinary materials with a positive susceptibility $\Re\suscd\geq 0$ (and $\Im\rho \leq 0$). The electric small (long-wavelength) limit is obtained by a radially inhomogeneous resistivity with values between 
%\begin{equation}
    $\rhot(0) \approx -\ka^2/6$
    and 
    $\rhot(a) \approx -\ka^2/15$.
%\end{equation}
This corresponds to a homogeneous susceptibility between
\begin{equation}
    \frac{6}{(k_0a)^2}
    \text{ at } r=0
    \text{ and }
    \frac{15}{(k_0a)^2}    
    \text{ at } r=a
\end{equation}
assuming losses $\rhort\ll k_0a$, which demonstrates that the TE mode requires a high contrast material for electrically small ($k_0a\ll 1$) objects as seen in Fig.~\ref{fig:SynthesisSph}.

Expanding the analytic solution for electrically small sizes for the electric dipole (TM) yields, an anisotropic reactivity of the form 
\begin{equation}
\Im\{\rhomt(\rv)\} 
\approx 
\Big(
\frac{1}{3}
+\frac{\kr^2}{5}
+\frac{13\kr^4}{700}
-\frac{11\ka^2}{30}
+\frac{23\ka^4}{210}
\Big)\Id 
+\left(\frac{\kr^2}{10}
+\frac{11\kr^4}{700}\right)
\rvh\rvh.
%-\frac{43(k_0a)^6}{2835}
	\label{eq:rhoTMsph}
\end{equation}
The synthesized reactivity is depicted with dashed curves for the $\thetavh\thetavh$ and $\phivh\phivh$ components and with dashed dotted curves for the $\rvh\rvh$ component in Fig.~\ref{fig:SynthesisSph}. The material is homogeneous and isotropic for $k_0a\ll 1$ but heterogeneous and anisotropic for larger $k_0a$.
The TM reactivity reduces to the classical plasmonic~\cite{Maier2007} resonance $\Im\rhot\approx 1/3$ in the long wavelength limit $k_0a\ll 1$ with a susceptibility having the expansion 
\begin{equation}
    \suscd(\rv)\approx
    -\left(3+\frac{33\ka^2}{10}-\frac{9\kr^2}{5}\right)\Id
    -\frac{9\kr^2}{10}\rvh\rvh
	\label{eq:chiTMsph}    
\end{equation}
for cases with negligible losses $\rhort\ll 1$. This positive reactivity $\Im\rhomrt\succ \M{0}$ and negative susceptibility $\Re\suscd\prec \M{0}$ are typical for small TM scatterers, \ie the dominant capacitance (electric) near-field need to be matched with an inductive material response. The asymptotic expansion~\eqref{eq:chiTMsph} can be compared with the results in~\cite{Nordebo+etal2019a,Tzarouchis+Sihvola2018} derived for homogeneous spheres by matching the asymptotic expansion coefficients.

\begin{figure}
    \centering
    \includegraphics[width=0.65\textwidth]{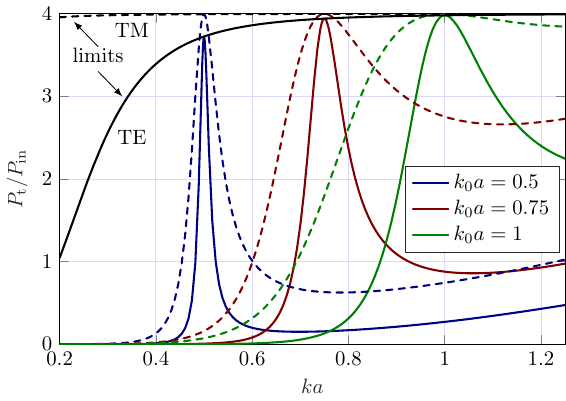}
    \caption{Frequency sweep of the extincted power for the synthesized material distributions designed for $k_0a\in\{0.5,0.75,1\}$ and losses $k\rhor/\eta_0=10^{-4}$. TM in dashed curves and TE in solid curves.}
    \label{fig:SynthesisSphereFreqSweep}
\end{figure}

Frequency ($ka$) sweeps for synthesized materials using resonance frequencies of $k_0a \in \{0.5, 0.75, 1\}$ are shown in Fig.~\ref{fig:SynthesisSphereFreqSweep}, assuming non-dispersive material models. The extinction power peaks at the resonance frequencies mentioned, reaching the limit~\eqref{eq:maxPtRradm}. TM cases exhibit higher peak values and broader half-power bandwidths. However, the bandwidth may be narrower when accounting for realistic material models with temporal dispersion for these plasmonic materials~\cite{Gustafsson+Sjoberg2010a}.

Results in Fig.~\ref{fig:SynthesisSph} and Fig.~\ref{fig:SynthesisSphereFreqSweep} as well as the asymptotic expansions~\eqref{eq:rhoTMsph} and~\eqref{eq:rhoTEsph} are given for the dipole ($l=1$) term. Synthesis for TE and TM dipole modes can also be combined, as they have perpendicular current distributions.
Higher order terms such quadrupole terms ($l=2$) are given by the closed form expressions in App.~\ref{S:SphWaves}.

\section{Numerical material synthesis}
Synthesis~\eqref{eq:PtRF2X} is performed numerically for arbitrary shaped regions. The numerical results are indistinguishable from the analytical for the spherical regions shown in Sec.~\ref{S:SynthesisSph}. The synthesized material distribution is typically anisotropic as for the TM case~\eqref{eq:rhoTMsph} and hence non-trivial to depict. 

\begin{figure}
    \centering
    \includegraphics[width=0.65\textwidth]{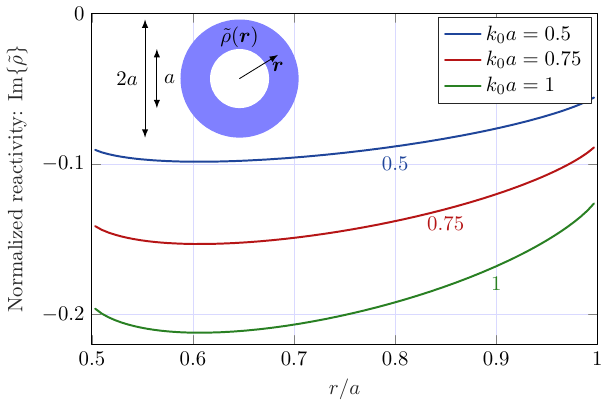}
    \caption{Synthesized surface reactivity for a planar loop structure with radius $a$ and width $a/2$ designed for $k_0a\in\{0.5,0.75,1\}$ and homogeneous losses $\rhort$. The material is synthesized for dominant TE radiation, \ie, azimuthally directed currents.}
    \label{fig:materialsynthesisBoRdisc}
\end{figure}

Numerical synthesis is here, for simplicity, demonstrated on a planar loop structure with a radius of $a$ and a strip width of $a/2$, as shown in the inset of Fig.~\ref{fig:materialsynthesisBoRdisc}. The loop has a uniform surface resistivity, $\rhort$, and the surface reactivity, $\Im\rhot(\rv) = \rhoit(\rv)$, is synthesized for maximal response for a $\phivh$-polarized far field related to the dominant TE mode. This corresponds to a $\phivh$-directed surface current and an isotropic resistivity similar to the synthesized material for the spherical TE mode in Fig.~\ref{fig:SynthesisSph}. The synthesized surface reactivity is depicted in Fig.~\ref{fig:materialsynthesisBoRdisc} for electrical sizes $k_0a \in \{0.5, 0.75, 1\}$. The reactivity is negative, similar to the TE sphere case, corresponding to a positive surface susceptibility. The reactivity depends on the radius $r$ and exhibits smaller values (greater contrast) for smaller sizes.

\section{Limits and synthesis for lossless materials}\label{S:lossless}
The results in Secs.~\ref{S:limits} and~\ref{S:Synthesis} are tight by allowing arbitrary reactive material parameters $\rhomi$. Synthesis shows that the resulting material often have a negative susceptibility $\Re\chi<0$, such as the plasmonic behavior for the TM case in~\eqref{eq:chiTMsph}  or has high contrast as for the TE case~\eqref{eq:rhoTEsph}. In many situations, we are restricted to use ordinary dielectric material with finite and positive susceptibility which can have low losses. Here, it is demonstrated how these constraints effect the performance.
Using conservation of complex power and starting with a single material~\cite{Gustafsson+etal2020} constructs the optimization problem~\eqref{eq:MaxPtZ}.
Lagrangian duality for a fixed excitation $\am$ reduces the optimization problem for maximal total interaction power to~\cite{Gustafsson+etal2020}
\begin{equation}
	P_{\mrm{tZ}}=\min_{\pm}\min_{\upsilon\in \Ds_{\pm}}
	\frac{1\pm\sqrt{1+\upsilon^2}}{4}\am^{\herm}\Um(\Rm+\upsilon\Xm)^{-1}\Um^{\herm}\am
\label{eq:MaxPtZ1}
\end{equation}
with the domains expressed in $\maxlambda=\max\lambda_n$ and $\minlambda=\min\lambda_n$
\begin{equation}
\begin{cases}    
    \Ds_{+} =[-\maxlambda^{-1},-\minlambda^{-1}],  \ \Ds_{-}=\emptyset, 
    & \Xm\ \T{indef.}\\
    \Ds_{+} =[-\maxlambda^{-1},\infty], \ 
    \Ds_{-} =[-\infty,-\minlambda^{-1}],  
    & \Xm\succ 0\\ 
    \Ds_{+} =[-\infty,-\minlambda^{-1}], \ 
     \Ds_{-} =[-\maxlambda^{-1},\infty],
     & \Xm \prec 0, 
\end{cases}
    \label{eq:MaxPtRange}
\end{equation}
where $\lambda_n$ denotes lossy characteristic modes (CM) satisfying~\cite{Gustafsson+etal2020}
\begin{equation}
    \Xm\Jm_n = \lambda_n\Rm\Jm_n
    \qtext{for }
    n=1,2,\ldots,N.
		\label{eq:CM_Jn}
\end{equation}
For lossy cases $\Rm\succ\Om$, the definiteness of $\Xm$ is determined by the characteristic values $\lambda_n$, \eg $\Xm\prec \Om\Leftrightarrow \maxlambda<0$. 

Maximizing~\eqref{eq:MaxPtZ1} over excitations $\M{a}$ results in the parametrized eigenvalue problem
\begin{equation}
	\max \frac{P_{\mrm{tZ}}}{P_{\mrm{in}}}
	=\max_{\am}8\frac{P_{\mrm{tZ}}}{|\am|^2}
	=2\min_{\pm}\min_{\upsilon\in \Ds_{\pm}}(1\pm\sqrt{1+\upsilon^2})\max\eig(\Um(\Rm+\upsilon\Xm)^{-1}\Um^{\herm}).
\label{eq:MaxPtZ2}
\end{equation}
This problem is solved numerically and is simplified by diagonalizing the matrices by the characteristic modes~\eqref{eq:CM_Jn}, where $\Qm^{\herm}\Rm\Qm=\Id$ is an identity matrix and $\Qm^{\herm}\Xm\Qm=\Lambdam$ is a diagonal matrix with the characteristic values $\lambda_n$ on the diagonal~\cite{Gustafsson+etal2020}.
 This simplifies the eigenvalue problem in~\eqref{eq:MaxPtZ2} to
\begin{equation}
    \max\eig(\Um\Qm(\Id+\upsilon\Lambdam)^{-1}\Qm^{\herm}\Um^{\herm}).
\label{eq:eq1}
\end{equation}
Here, $\Um\Qm$ is interpreted as the radiated field of the characteristic current, suggesting an efficiency $\Qm^{\herm}\Rmr\Qm\prec\Id$ for lossy cases $\Rm\succ\Om$. This efficiency suppresses contributions from weakly radiating CM modes. 

To analyze objects with negligible or no losses, we start to analyze the behavior of the characteristic modes in the limit of vanishing losses.
Consider a MoM system $\Zm=\Zm_0+\rho\Psim$ with complex resistivity $\rho=\rhor+\ju\rhoi$ and Gramian matrix $\Psim$.
The derivative of non-degenerate lossy CM~\eqref{eq:CM_Jn} $\lambda_n=\eig(\Xm,\Rm)$ with respect to the imaginary part $\rhoi$ is
\begin{equation}
  \partder{\lambda_n}{\rhoi} = \frac{\Jm^{\herm}\big(\partder{\Xm}{\rhoi}-\lambda_n\partder{\Rm}{\rhoi}\big)\Jm_n}{\Jm_n^{\herm}\Rm\Jm_n}
  =\frac{\Jm_n^{\herm}\Psim\Jm_n}{\Jm_n^{\herm}\Rm\Jm_n}>0
\label{eq:lambdaprim}
\end{equation}
as the resistance matrix $\M{R}$ is independent of $\rhoi$.
Here, we observe that $\lambda_n(\rhoi)$ increases monotonically with $\rhoi$ for the lossy case $\Rm\succ\Om$. For lossless CM $\lambda_n(\rhoi)$ increases except for non-radiating (inner resonances) where $\Rm=\Rmr=\Om$. Degenerate eigenvalues are related to geometrical symmetries and can be decomposed into separate subspaces with non-degenerate eigenvalues~\cite{Wigner+Neumann1929,Schab+Bernhard2016}.
The characteristic modes are not ordered but instead tracked as commonly done for CM. 
Monotonicity of $\lambda_n$ in $\rhoi$~\eqref{eq:lambdaprim} is also valid for inhomogeneous regions and local increase of $\rhoi(\rv)$ or more generally $\rhomi(\rv)$. Moreover, sensitivity with respect to $\rhor$  
\begin{equation}
  \partder{\lambda_n}{\rhor} = -\lambda_n\frac{\Jm_n^{\herm}\Psim\Jm_n}{\Jm_n^{\herm}\Rm\Jm_n}	
	\ \begin{cases}
		<0 & \text{for }\lambda_n >0 \\
		>0 & \text{for }\lambda_n <0 \\		
	\end{cases}
\label{eq:eq2}
\end{equation} 
shows that $\lambda_n$ approach zero with increasing losses $\rhor$.

\begin{figure}%
\includegraphics[width=0.49\textwidth]{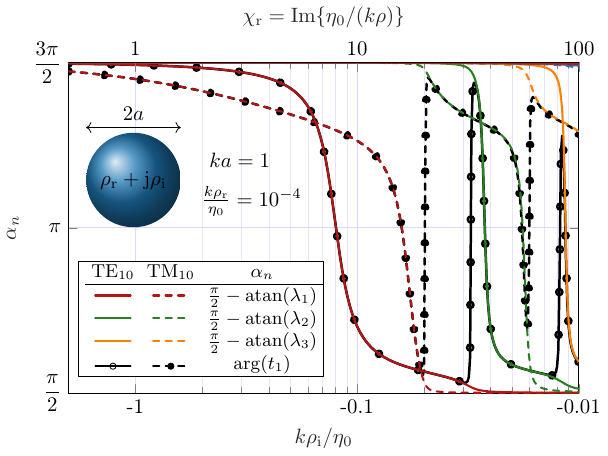}%
\includegraphics[width=0.49\textwidth]{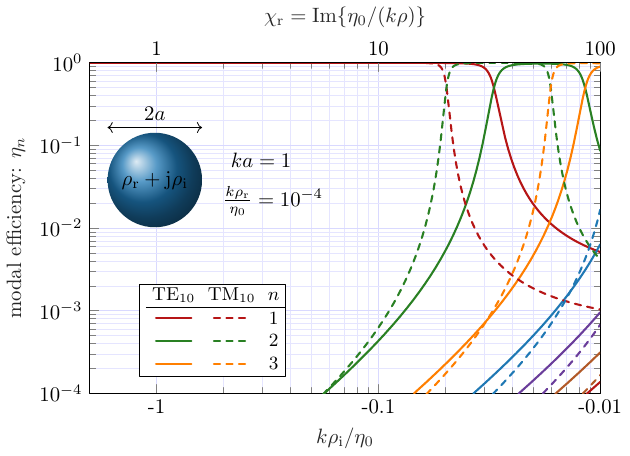}

\caption{Characteristic angles $\alpha_n$ (top) and efficiencies (bottom) for a spherical region with electrical size $ka=1$, resistivity $\rho=\rhor+\ju\rhoi$ with $k\rhor=10^{-4}\eta_0$, azimuthal index $m=0$, and dipole pattern $l=1$. Horizontal axes $k\rhoi/\eta_0$ on bottom and $\Re\{\suscd\}$ on top. 
Characteristic eigenvalues from MoM matrices $\lambda_n=\eig(\Xm,\Rm)$ from~\eqref{eq:CM_Jn} and scattering formulation (Mie series) with transition matrix eigenvalues $t_n$~\cite{Gustafsson+etal2022a}.}%
\label{fig:CMlossy}%
\end{figure}

We analyze a spherical region to compare lossy characteristic modes (CMs) determined from the MoM system matrices~\eqref{eq:CM_Jn} with those derived from scattering problems (Mie series)~\cite{Gustafsson+etal2022a}. The spherical symmetry is used to simplify the problem to the dipole ($l=1$) mode, as depicted in Fig.~\ref{fig:CMlossy}. The monotonic increase in the CM eigenvalues $\lambda_n$ corresponds to the monotonic decrease in the characteristic angles $\alpha_n = \pi/2 - \atan(\lambda_n)$, as seen in the figure. 

There are infinitely many CMs as $P \to \infty$ from the eigenvalue problem~\eqref{eq:CM_Jn}, with all characteristic angles $\alpha_n$ starting at $3\pi/2$ (or $\lambda_n = -\infty$) in free space ($\rhoi = -\infty$) and approaching $\pi/2$ as $\rhoi \to \infty$. This behavior is radically different from that of scattering-based CMs~\cite{Gustafsson+etal2022a}, illustrated by the two $\arg(t_1)$ graphs. Only two scattering modes—represented by the TE and TM modes—exist for the depicted dipole ($l=1$) modes. These modes closely follow the $\eig(\Xm, \Rm)$ modes downward until approximately $\pi/2$, where they quickly turn upward toward $3\pi/2$ before aligning with the next $\eig(\Xm, \Rm)$ mode downward. 
This difference is also evident from the modal efficiency in Fig.~\ref{fig:CMlossy}, which shows the efficiency of the characteristic modes using~\eqref{eq:CM_Jn}. The first ($n=1$) modes exhibit high efficiency (close to 1) up to a cutoff point, where the efficiency rapidly decreases. At the cutoff, the efficiency of the next modes ($n=2$) rapidly increases. The corresponding efficiencies of the scattering TE and TM modes remain consistently close to unity.

This difference is negligible in many classical CM applications, such as summation formulas and expansions of currents and far fields~\cite{Chen+Wang2015}. However, the difference is fundamental for the bounds analyzed here, where the domains for the Lagrange parameters~\eqref{eq:MaxPtRange} depend on the definiteness of $\Xm$ (signs of $\maxlambda$ and $\minlambda$). Bounds for lossless cases solved using CM must hence be treated in the limit of vanishing losses, where CM modes approaching $\pi/2$ stay at $\pi/2$, and new modes start from $3\pi/2$. 
For a given resistivity, CMs are hence given by the union of lossless CMs and a set of non-radiating modes with $\alpha_n=\pi/2$ (or equivalently $\lambda_n=+\infty$). These modes can also be useful for expansions of the current inside the region.

Lossless objects are described by a real-valued permittivity or susceptibility corresponding to an imaginary resistivity, \ie $\rhom=\ju\rhomi$ with $\rhomi = -\frac{\eta_0}{k}\suscd_{\mrm{r}}^{-1}$,
where it is noted that a small positive susceptibility corresponds to a large negative imaginary resistivity. 

\begin{figure}
\centering
\includegraphics[height=0.35\columnwidth]{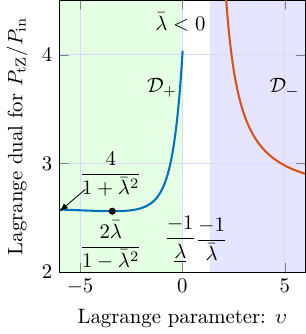}%
\hspace{2mm}
\includegraphics[height=0.35\columnwidth]{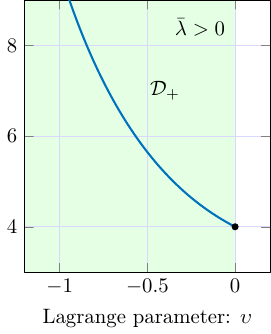}
\caption{Solution of~\eqref{eq:PtZlosslessCM} illustrated for a negative definite $-1<\maxlambda<0$ case (left) and an indefinite $\maxlambda>0$ case (right) with $\minlambda=-100$. The negative definite $\maxlambda<0$ case has a stationary point at $2\maxlambda/(1-\maxlambda^2)$ with value $P_\T{tZ}/P_\T{in}=4/(1+\maxlambda^2)$. The indefinite case $\maxlambda>0$ has $P_\T{tZ}/P_\T{in}\to 4$ as $\minlambda\to-\infty$ at the boundary point $\upsilon=0$.}
\label{fig:PtZCMdual}
\end{figure}

Expansion of~\eqref{eq:MaxPtZ2} in lossy characteristic modes~\eqref{eq:CM_Jn} in the limit of lossless materials reduces the optimization problem~\eqref{eq:MaxPtZ2} to
\begin{equation}
	\max\frac{P_{\mrm{tZ}}}{P_{\mrm{in}}}
	=2\min_{\pm}\min_{\upsilon\in \Ds_{\pm}}
	\max_n\frac{1\pm\sqrt{1+\upsilon^2}}{\lambda_n\upsilon+1}.
\label{eq:PtZlosslessCM}
\end{equation}
Solution of~\eqref{eq:PtZlosslessCM} depends on the properties of the reactance matrix~\eqref{eq:MaxPtRange} determined by the largest $\maxlambda$ and smallest  $\minlambda$ characteristic eigenvalues. Starting with the negative definite case $\maxlambda<0$ together with $\minlambda\to-\infty$ as typical for dielectric objects, \cf Fig.~\ref{fig:CMlossy}. The Lagrange dual~\eqref{eq:PtZlosslessCM} is depicted in Fig.~\ref{fig:PtZCMdual} for $\maxlambda=-0.75$ having a stationary point $\upsilon=2\maxlambda/(1-\maxlambda^2)\approx -3.4$ resulting in the bound $P_\T{tZ}/P_\T{in}\leq 4/(1+\maxlambda^2)\approx  2.6$. The stationary point moves to the $\Ds_-$ region for $\maxlambda<-1$ with the same result $4/(1+\maxlambda^2)$. The indefinite case with $\maxlambda>0$ and $\minlambda\approx-\infty$ is depicted in Fig.~\ref{fig:PtZCMdual} for $\maxlambda=0.5$. The minimum value gives the limit $P_\T{tZ}/P_\T{in}\leq 4$ for all $\maxlambda>0$. For these dielectric objects we have the limit
\begin{equation}
	\frac{P_{\mrm{t}}}{P_\T{in}}
	\leq \begin{cases}
		\dfrac{4}{1+\maxlambda^2} & \text{for }\maxlambda<0\\
		4 & \text{for } \maxlambda\geq 0. 
	\end{cases}
\label{eq:PtZlosslessCMres}
\end{equation}
Here, we note that scattering situations with negative characteristic modes $\maxlambda=\max\lambda_n\leq 0$ for maximum contrast material are optimal and the maximal value corresponds to the squared maximal modal significance $4/(1+\lambda_n^2)=4|t_n|^2$. This is the normalized scattered power of an object composed by $\rhomi$ expressed in characteristic eigenvalues $\lambda_n$~\cite{Gustafsson+etal2022a}.

The limit in~\eqref{eq:PtZlosslessCMres} is derived using the equality constraint $\Jm^{\herm}\Zm\Jm=\Jm^{\herm}\Vm$ from~\eqref{eq:MaxPtZ} in the case of lossless dielectric materials with a positive susceptibility $\suscd \succ \M{0}$. This limit remains the same when the equality constraint is relaxed to an inequality, permitting susceptibilities ranging from $\suscd$ down to $\M{0}$ (vacuum). This relaxation can also be interpreted through homogenization theory: a material that combines both $\suscd$ and vacuum results in a composite with a susceptibility value between $\suscd$ and $\M{0}$~\cite{Milton2002}.

The explicit construction using characteristic modes shows that the bound is tight for $\maxlambda\leq 0$. For $\maxlambda\geq 0$, we note that the characteristic values are monotonically increasing in $\rhoi$~\eqref{eq:lambdaprim}.
The eigenvalues are also continuous in $\rhoi$, so deceasing a negative $\rhoi$ decreases $\lambda_n$. This corresponds to decreasing $\suscd$ until $\maxlambda=0$ and a tight bound for this case. 

Perfect electric conductors (PEC) can be interpreted as a high contrast material, and the maximum characteristic eigenvalue is always positive for PEC objects. The bound~\eqref{eq:PtZlosslessCMres} is tight for PEC objects with $P_\T{t}/P_\T{in}=4$ with the interpretation that arbitrary inductive or capacitive loading can be synthesized in a PEC structure. 
  
\section{Conclusions}\label{S:Conclusions}
This study presents a comprehensive approach to material synthesis for maximizing extinction, scattering, and scattering of arbitrarily shaped objects under optimal far-field illumination. The proposed solutions leverage radiation modes to achieve minimal resistivity in materials and employ characteristic modes to optimize contrast in lossless dielectric materials. The findings demonstrate scenarios with stringent physical limits and highlight instances where optimal currents enable efficient material synthesis. Numerical synthesis methods are applied to objects of arbitrary shapes, while analytical synthesis is used for spherical regions. The obtained normalized limits for extinct, scattered, and absorbed powers can also be interpreted as limits on the transition matrix eigenvalues. 

\appendix
\section{Spherical waves}\label{S:SphWaves}
The scattered electric and magnetic fields are expressed in the contrast sources by the Green's dyadic $\M{G}$ which is expanded in regular $\uop^{(1)}_{\sphi}$ and outgoing $\uop^{(4)}_{\sphi}$ spherical waves as~\cite{Hansen1988,Kristensson2016}
\begin{multline}
  \Ev_\T{s}(\rv_1) 
	=-\ju k\eta_0\int_{\reg} \M{G}(\rv_1-\rv_2)\cdot\Jv(\rv_2)\diffV_2\\
  =\eta_0 k^2 \sum_{\sphi}\uop^{\mrm{(4)}}_{\sphi}(k\rv_1) \int_{\reg}\uop^{\mrm{(1)}}_{\sphi}(k\rv_2)\cdot\Jv(\rv_2)\diffV_2  
\label{eq:eq3}
\end{multline}
for $\rv_1$ outside a sphere circumscribing the region $\reg$.
Define the spherical wave expansion matrix $\Um$ with elements~\cite{Tayli+etal2018}
\begin{equation}
  U_{\sphi p} = k\sqrt{\eta_0}\int_\reg\uop^{{(1)}}_{\sphi}(k\rv)\cdot\basv_p(\rv)\diffV
\label{eq:Smatrix}
\end{equation}
to express the scattered spherical mode coefficients from the current $\Jm$ as $-\Um\Jm$ and the total radiated power as $P_{\mrm{r}} = \frac{1}{2}|\Um\Jm|^2 = \frac{1}{2}\Jm^{\herm}\Rmr\Jm$.
Spherical waves $\uv_n^{(p)}$ have radial dependence expressed in radial functions $\Rop_{\tau l}^{(p)}(kr)$~\cite{Hansen1988} of order $l$:
\begin{equation}
	\Rop_{\tau l}^{(p)}(\kr)
	=\begin{cases}
		z_l^{(p)}(\kr) & \tau=1\\
		\displaystyle{\frac{1}{\kr}\partder{(\kr z_l^{(p)}(\kr))}{\kr}} & \tau=2\\
		\sqrt{l(l+1)}z_l^{(p)}(\kr)/\kr & \tau=3.
	\end{cases}
\end{equation}
For regular waves ($p=1$) $z_l^{(1)}=\T{j}_l$ is a spherical Bessel function, irregular waves ($p=2$) $z_l^{(2)}=\T{n}_l$ is a spherical Neumann function, and outgoing waves ($p=4$) $z_l^{(4)}=\T{h}^{(2)}_l$ is an outgoing spherical Hankel function.

Radiation modes for homogeneous spherical (ball) region with normalized resistivity $\rhort$ is expressed in spherical waves as~\cite{Gustafsson+etal2020}
\begin{equation}
		\radm_n = \frac{k^3}{\rhort}\int_{r\leq a}
		|\uv_{\nu}(kr)|^2\diffV
		=\frac{k^3 a^3}{2\rhort}
		\left(
		(\Rop_{1,l}^{(1)})^2 - \Rop_{1,l-1}^{(1)}\Rop_{1,l+1}^{(1)} + \frac{2}{ka}\Rop_{1,l}^{(1)}R_{2,l}^{(1)}\delta_{\tau,2}
		\right).
	\label{eq:sphrad}
\end{equation}
For TE ($\tau=1$) radiation modes with currents of the form $\Rop_{1,l}^{(1)}(\kr)\Yv_{1lm}(\rvh)$ with spherical harmonics $\Yv_{1lm}$~\cite{Hansen1988,Kristensson2016} generate~\eqref{eq:Xsynthesis} an isotropic radially dependent reactivity $\rhoi(r)$ 
\begin{equation}
\rhoi = \frac{\Rop_{1l}^{(2)}(kr)}{\Rop_{1l}^{(1)}(kr)}\int_0^{kr} 
|x\Rop_{1l}^{(1)}|^2 \diff x
+\int_{kr}^{ka} x^2\Rop_{1l}^{(1)}\Rop_{1l}^{(2)} \diff x,
\label{eq:SphBesselTE}
\end{equation}
These integrals have closed form expressions based on the identity 
\begin{equation}
	\int x^2 z_m(x) y_m(x)\diff x 
=\frac{x^3}{4}\big(2z_m y_m-z_{m-1} y_{m+1}-z_{m+1} y_{m-1}\big)
\label{eq:SphBesselId}
\end{equation}
with $z_m,y_m$ denoting arbitrary spherical Bessel, Neumann, or Hankel functions. For the dipole term ($l=1$)~\eqref{eq:SphBesselTE} using~\eqref{eq:SphBesselId} simplifies to~\eqref{eq:rhoTEsph}.

The expressions are more involved for the TM ($\tau=1$) case having an anisotropic radially dependent reactivity $\rhomi(r)$. To simplify notation, introduce the quantity 
\begin{equation}
	h_{pq}(\kr,\ka) 
		=\left[x^2\Rop^{(p)}_{1l}\Rop_{2l}^{(q)\ast}\right]_{\kr}^{\ka}
	+\int_{\kr}^{\ka}x^2\Rop_{1l}^{(p)}\Rop_{1l}^{(q)\ast} \diff x
\label{eq:hfunk}
\end{equation}	
expressible in Bessel and Neumann functions using the identity~\eqref{eq:SphBesselId}. The transverse part $\UV{\theta}\UV{\theta}+\UV{\phi}\UV{\phi}$ of the reactivity is
\begin{equation}
		\rhoit{}_\T{t}(r) = \frac{R_{2l}^{(2)}(kr)}{R_{2l}^{(1)}(kr)}h_{11}(0,kr)
		+h_{12}(kr,ka)
	\label{eq:eq4}
\end{equation}	
and radial part $\UV{r}\UV{r}$ 
\begin{equation}
\rhoit{}_\T{r}(r) = \frac{R_{3l}^{(2)}(kr)}{R_{3l}^{(1)}(kr)}h_{11}(0,kr)
+h_{12}(kr,ka)+1.
\label{eq:eq5}
\end{equation}		
These expressions are valid for all orders, and the small size asymptotic for the dipole term is given in~\eqref{eq:rhoTMsph}.

%\section*{References}

%\bibliographystyle{IEEEtran}
%\bibliography{total}

\begin{thebibliography}{10}
\providecommand{\url}[1]{#1}
\csname url@samestyle\endcsname
\providecommand{\newblock}{\relax}
\providecommand{\bibinfo}[2]{#2}
\providecommand{\BIBentrySTDinterwordspacing}{\spaceskip=0pt\relax}
\providecommand{\BIBentryALTinterwordstretchfactor}{4}
\providecommand{\BIBentryALTinterwordspacing}{\spaceskip=\fontdimen2\font plus
\BIBentryALTinterwordstretchfactor\fontdimen3\font minus
  \fontdimen4\font\relax}
\providecommand{\BIBforeignlanguage}[2]{{%
\expandafter\ifx\csname l@#1\endcsname\relax
\typeout{** WARNING: IEEEtran.bst: No hyphenation pattern has been}%
\typeout{** loaded for the language `#1'. Using the pattern for}%
\typeout{** the default language instead.}%
\else
\language=\csname l@#1\endcsname
\fi
#2}}
\providecommand{\BIBdecl}{\relax}
\BIBdecl

\bibitem{Molesky+etal2018}
S.~Molesky, Z.~Lin, A.~Y. Piggott, W.~Jin, J.~Vuckovi{\'c}, and A.~W.
  Rodriguez, ``Inverse design in nanophotonics,'' \emph{Nature Photonics},
  vol.~12, no.~11, pp. 659--670, 2018.

\bibitem{Koziel+Ogurtsov2014}
S.~Koziel and S.~Ogurtsov, \emph{Antenna Design by Simulation-Driven
  Optimization}.\hskip 1em plus 0.5em minus 0.4em\relax Springer, 2014.

\bibitem{Rahmat-Samii+Michielssen1999}
Y.~Rahmat-Samii and E.~Michielssen, \emph{Electromagnetic Optimization by
  Genetic Algorithms}, ser. Wiley Series in Microwave and Optical
  Engineering.\hskip 1em plus 0.5em minus 0.4em\relax John Wiley \& Sons, 1999.

\bibitem{Gustafsson+Nordebo2013}
M.~Gustafsson and S.~Nordebo, ``Optimal antenna currents for {Q},
  superdirectivity, and radiation patterns using convex optimization,''
  \emph{IEEE Trans. Antennas Propag.}, vol.~61, no.~3, pp. 1109--1118, 2013.

\bibitem{Gustafsson+etal2020}
M.~Gustafsson, K.~Schab, L.~Jelinek, and M.~Capek, ``Upper bounds on absorption
  and scattering,'' \emph{New Journal of Physics}, vol.~22, no. 073013, 2020.

\bibitem{Kuang+Miller2020}
Z.~Kuang and O.~D. Miller, ``Computational bounds to light-matter interactions
  via local conservation laws,'' \emph{Phys. Rev. Lett.}, vol. 125, no. 263607,
  2020.

\bibitem{Chao+etal2022}
P.~Chao, B.~Strekha, R.~Kuate~Defo, S.~Molesky, and A.~W. Rodriguez, ``Physical
  limits in electromagnetism,'' \emph{Nature Reviews Physics}, vol.~4, no.~8,
  pp. 543--559, 2022.

\bibitem{Kuang+etal2020}
Z.~Kuang, L.~Zhang, and O.~D. Miller, ``Maximal single-frequency
  electromagnetic response,'' \emph{Optica}, vol.~7, pp. 1746--1757, 2020.

\bibitem{Capek+etal2019b}
M.~Capek, L.~Jelinek, K.~Schab, M.~Gustafsson, B.~L.~G. Jonsson, F.~Ferrero,
  and C.~Ehrenborg, ``Optimal planar electric dipole antennas,'' \emph{IEEE
  Antennas Propag. Mag.}, vol.~61, no.~4, pp. 19--29, 2019.

\bibitem{Gustafsson+etal2009a}
M.~Gustafsson, C.~Sohl, and G.~Kristensson, ``Illustrations of new physical
  bounds on linearly polarized antennas,'' \emph{IEEE Trans. Antennas Propag.},
  vol.~57, no.~5, pp. 1319--1327, May 2009.

\bibitem{Nel+etal2023}
B.~A.~P. Nel, A.~K. Skrivervik, and M.~Gustafsson, ``Q-factor bounds for
  microstrip patch antennas,'' \emph{IEEE Trans. Antennas Propag.}, vol.~71,
  no.~4, pp. 3430--3440, 2023.

\bibitem{Gustafsson+Capek2020}
M.~Gustafsson and M.~Capek, ``Physical bounds on antennas with feed
  constraints,'' in \emph{14th European Conference on Antennas and Propagation
  (EuCAP)}, 2020.

\bibitem{Schab2016}
K.~R. Schab, ``Modal analysis of radiation and energy storage mechanisms on
  conducting scatterers,'' Ph.D. dissertation, University of Illinois at
  Urbana-Champaign, 2016.

\bibitem{Miller+etal2016}
O.~D. Miller, A.~G. Polimeridis, M.~H. Reid, C.~W. Hsu, B.~G. DeLacy, J.~D.
  Joannopoulos, M.~Solja{\v{c}}i{\'c}, and S.~G. Johnson, ``Fundamental limits
  to optical response in absorptive systems,'' \emph{Optics express}, vol.~24,
  no.~4, pp. 3329--3364, 2016.

\bibitem{Garbacz+Turpin1971}
R.~J. Garbacz and R.~H. Turpin, ``A generalized expansion for radiated and
  scattered fields,'' \emph{IEEE Trans. Antennas Propag.}, vol.~19, no.~3, pp.
  348--358, 1971.

\bibitem{Harrington+Mautz1971}
R.~F. Harrington and J.~R. Mautz, ``Theory of characteristic modes for
  conducting bodies,'' \emph{IEEE Trans. Antennas Propag.}, vol.~19, no.~5, pp.
  622--628, 1971.

\bibitem{Gustafsson+etal2022a}
M.~Gustafsson, L.~Jelinek, K.~Schab, and M.~Capek, ``Unified theory of
  characteristic modes: Part {I}--fundamentals,'' \emph{IEEE Trans. Antennas
  Propag.}, vol.~70, no.~12, pp. 11\,801--11\,813, 2022.

\bibitem{Kristensson2016}
G.~Kristensson, \emph{Scattering of Electromagnetic Waves by Obstacles}.\hskip
  1em plus 0.5em minus 0.4em\relax Edison, NJ: SciTech Publishing, an imprint
  of the IET, 2016.

\bibitem{Hansen1988}
J.~E. Hansen, Ed., \emph{Spherical Near-Field Antenna Measurements}, ser. {IEE}
  electromagnetic waves series.\hskip 1em plus 0.5em minus 0.4em\relax
  Stevenage, UK: Peter Peregrinus Ltd., 1988, no.~26.

\bibitem{Capek+etal2023a}
M.~Capek, J.~Lundgren, M.~Gustafsson, K.~Schab, and L.~Jelinek,
  ``Characteristic mode decomposition using the scattering dyadic in arbitrary
  full-wave solvers,'' \emph{IEEE Trans. Antennas Propag.}, vol.~71, no.~1, pp.
  830--839, 2023.

\bibitem{Gustafsson+etal2015b}
M.~Gustafsson, D.~Tayli, and M.~Cismasu, \emph{Physical bounds of
  antennas}.\hskip 1em plus 0.5em minus 0.4em\relax Sprin\-ger-Verlag, 2015,
  pp. 197--233.

\bibitem{Harrington1968}
R.~F. Harrington, \emph{Field Computation by Moment Methods}.\hskip 1em plus
  0.5em minus 0.4em\relax New York, NY: Macmillan, 1968.

\bibitem{Milton2002}
G.~W. Milton, \emph{The Theory of Composites}.\hskip 1em plus 0.5em minus
  0.4em\relax Cambridge: Cambridge University Press, 2002.

\bibitem{Boyd+Vandenberghe2004}
S.~P. Boyd and L.~Vandenberghe, \emph{{Convex Optimization}}.\hskip 1em plus
  0.5em minus 0.4em\relax Cambridge Univ. Pr., 2004.

\bibitem{Gustafsson+Capek2019}
M.~Gustafsson and M.~Capek, ``Maximum gain, effective area, and directivity,''
  \emph{IEEE Trans. Antennas Propag.}, vol.~67, no.~8, pp. 5282--5293, 2019.

\bibitem{Ehrenborg+Gustafsson2020}
C.~Ehrenborg and M.~Gustafsson, ``Physical bounds and radiation modes for
  {MIMO} antennas,'' \emph{IEEE Trans. Antennas Propag.}, vol.~68, no.~6, pp.
  4302--4311, 2020.

\bibitem{Gustafsson2024a}
M.~Gustafsson, ``Degrees of freedom for radiating systems,'' \emph{arXiv
  preprint arXiv:2404.08976}, 2024.

\bibitem{Peierls1979}
R.~E. Peierls, \emph{Surprises in Theoretical Physics}.\hskip 1em plus 0.5em
  minus 0.4em\relax Princeton University Press, 1979.

\bibitem{Polimeridis+etal2014}
A.~G. Polimeridis, J.~F. Villena, L.~Daniel, and J.~K. White, ``Stable
  {FFT-JVIE} solvers for fast analysis of highly inhomogeneous dielectric
  objects,'' \emph{Journal of Computational Physics}, vol. 269, pp. 280--296,
  2014.

\bibitem{Maier2007}
S.~A. Maier, \emph{Plasmonics: Fundamentals and Applications}.\hskip 1em plus
  0.5em minus 0.4em\relax Berlin: Sprin\-ger-Verlag, 2007.

\bibitem{Nordebo+etal2019a}
S.~Nordebo, G.~Kristensson, M.~Mirmoosa, and S.~Tretyakov, ``Optimal plasmonic
  multipole resonances of a sphere in lossy media,'' \emph{Phys. Rev. B},
  vol.~99, no.~5, p. 054301, 2019.

\bibitem{Tzarouchis+Sihvola2018}
D.~Tzarouchis and A.~Sihvola, ``Light scattering by a dielectric sphere:
  perspectives on the mie resonances,'' \emph{Applied Sciences}, vol.~8, no.~2,
  p. 184, 2018.

\bibitem{Gustafsson+Sjoberg2010a}
M.~Gustafsson and D.~Sj\"{o}berg, ``Sum rules and physical bounds on passive
  metamaterials,'' \emph{New Journal of Physics}, vol.~12, no. 043046, pp.
  1--18, 2010.

\bibitem{Wigner+Neumann1929}
E.~Wigner and J.~Von~Neumann, ``On the behaviour of eigenvalues in adiabatic
  processes,'' \emph{Phys. Z}, vol.~30, no. 467, 1929.

\bibitem{Schab+Bernhard2016}
K.~Schab and J.~Bernhard, ``A group theory rule for predicting eigenvalue
  crossings in characteristic mode analyses,'' \emph{IEEE Antennas Wirel Propag
  Lett}, vol.~16, pp. 944--947, 2016.

\bibitem{Chen+Wang2015}
Y.~Chen and C.-F. Wang, \emph{Characteristic Modes: Theory and Applications in
  Antenna Engineering}.\hskip 1em plus 0.5em minus 0.4em\relax John Wiley \&
  Sons, 2015.

\bibitem{Tayli+etal2018}
D.~Tayli, M.~Capek, L.~Akrou, V.~Losenicky, L.~Jelinek, and M.~Gustafsson,
  ``Accurate and efficient evaluation of characteristic modes,'' \emph{IEEE
  Trans. Antennas Propag.}, vol.~66, no.~12, pp. 7066--7075, 2018.

\end{thebibliography}

\end{document}